\documentclass[a4paper]{article}
\usepackage[margin=1.5in]{geometry}
\usepackage[english]{babel}
\usepackage{newtxtext,newtxmath}
\usepackage{amsmath}
\usepackage{graphicx}
\usepackage{authblk}
\usepackage{float}
\usepackage[numbers]{natbib}
\usepackage[font=small]{caption}

\usepackage{titlesec}
\titleformat{\section}
  {\normalfont\normalsize}
  {\thesection}{1em}{}
\titleformat{\subsection}
  {\normalfont\small}
  {\thesubsection}{1em}{}
\titleformat{\subsubsection}
  {\normalfont\small}
  {\thesubsubsection}{1em}{}

\date{}

\begin{document}

\title{\bf \large Behavioral and Topological Heterogeneities in Network Versions of Schelling’s Segregation Model}
\author{\normalsize Will Deter$^{1*}$ and Hiroki Sayama$^{2}$\\
\small$^{1,2}$ Binghamton Center of Complex Systems, School of Systems Science and Industrial Engineering, Binghamton University\\
\small$^*$ wdeter1@binghamton.edu}

\bibliographystyle{unsrtnat}

\maketitle

\section{Abstract}

Agent-based models of residential segregation have been of persistent interest to various research communities since their origin with James Sakoda and popularization by Thomas Schelling. Frequently, these models have sought to elucidate the extent to which the collective dynamics of individual preferences may cause segregation to emerge. This open question has sustained relevance in U.S. jurisprudence. Previous investigation of heterogeneity of behaviors (preferences) has shown reductions in segregation. Meanwhile, previous investigation of heterogeneity of social network topologies has shown no significant impact to observed segregation levels.  In the present study, we examined the effects of the concurrent presence of both behavioral and topological heterogeneities in network segregation models. Simulations were conducted using both homogeneous and heterogeneous preference models on 2D lattices with varied levels of densification to create topological heterogeneities (i.e., clusters, hubs). Results show a richer variety of outcomes, including novel differences in resultant segregation levels and hub composition. Notably, with concurrent increased representations of heterogeneous preferences and heterogeneous topologies, reduced levels of segregation emerge. Simultaneously, we observe a novel dynamic of segregation between tolerance levels as highly tolerant nodes take residence in dense areas and push intolerant nodes to sparse areas mimicking the urban-rural divide.

\section{Introduction}
Residential segregation is a persistent topic of discussion in the social sciences whose investigation is frequently enriched by computational models, particularly agent-based models (ABMs) with network structure \cite{SakodaJamesM.1971Tcmo,schelling1971dynamic,xie2012modeling,gandica2016can,BRUCHElizabethE2006Ncan,clark2008understanding,ClarkW.A.V.1991RPaN,fossett2006ethnic,GargiuloFloriana2015UsfS,zhang2004dynamic,zhang2004residential,AbellaDavid2022AeiS,FagioloGiorgio2007Sin,GambettaDaniele2023Mcis,VinkovićDejan2006paot,MantzarisAlexanderV.2020Iamv}. Residential segregation appears to be a robust and resilient phenomenon with many possible contributing factors at multiple scales \cite{ClarkW.A.V.1991RPaN}. Several authors have demonstrated significant interplay between racial segregation and wealth and income inequality \cite{gandica2016can,BRUCHElizabethE2006Ncan, MasseyDouglasS.1990AASa,ReardonSeanF.2011IIaI}. Others have developed supporting theoretical and simulation models to understand this relationship \cite{benard2007wealth,gauvin2013modeling, sahasranaman2016dynamics,sahasranaman2017cooperative,HartingPhilipp2020RsTr,Bonakdar2023,vicario2024dynamic}. Identifying and understanding factors and behaviors which contribute to residential segregation has substantial sociopolitical significance.  A legal debate over the causes of segregation emerged with the Brown v. Board of Education decision in 1954. Whether segregation is a de jure or de facto phenomenon remains unresolved. The government's long-standing view is that it should restrict its interventions to cases where de jure segregation is evident.

Of the two traditions of segregation theory identified by Fossett \cite{fossett2006ethnic}: “social distance” and “individual preferences,” this work focuses on the latter which asserts that segregation is an emergent property arising from the collective dynamics of individuals and their choices in mostly-free housing markets.  Proponents of the de facto perspective cite the collective dynamics of individual preferences as the genesis of de facto segregation \cite{FrankenbergErica2018DFST}. Given the current state of related jurisprudence, it is clearly important to understand the extent to which de facto segregation might (or might not) emerge as the result of the collective dynamics of individual preferences.

Recent works in complex systems and network sciences \cite{sayama2020beyond,YamanoiJunichi2013Pcif,BullockSeth2023AHME,Sánchez-PuigFernanda2022Hec} have highlighted the importance of heterogeneity in network models of social problems, showing clearly that certain emergent system-level phenomena may only arise with sufficient heterogeneity of agent behaviors and/or characteristics.  A review of the literature identified many investigations of component heterogeneity in network models of segregation.  Notably, Xie and Zhou \cite{xie2012modeling} and Gandica, Gargiulo, \& Carletti \cite{gandica2016can} examined heterogeneous preferences and heterogeneous topology, respectively. The former obtained reductions in resultant segregation when heterogeneous preferences were represented while the latter showed no substantial reduction in segregation when heterogeneous topology was represented. No previous studies of combined dimensions of heterogeneity were identified. 
Recognizing the possibility that multiple dimensions of heterogeneity may be required to observe additional phenomena, this study attempts to examine the impact of the combination of behavioral and topological heterogeneities in network models of residential segregation. A baseline model is constructed following Xie \& Zhou’s model \cite{xie2012modeling}. The model is then elaborated in two ways: first to introduce heterogeneity of preferences and second to introduce heterogeneity of topology. The results of these implementations are illustrated, observables are characterized, and simulations are enumerated. Finally, results, novel behaviors, and opportunities for future work are discussed.

\section{Related Literature}

\subsection{Sakoda and Schelling}
In 1971, Sakoda \cite{SakodaJamesM.1971Tcmo} and then Schelling \cite{schelling1971dynamic}, proposed lattice-based models of social interaction and, more specifically, segregation. These models can be considered some of the earliest instances of agent-based models (ABMs). While Sakoda’s work faded into relative obscurity, Schelling’s gained great renown \cite{HegselmannRainer2017TCSa}. Importantly, these works demonstrated the emergence of macro-level (population-level) phenomena that resulted from micro-level (individual) interactions with neighbors. Both showed that segregation was one of the macro-level properties which could emerge.
Schelling’s investigation of segregation was more extensive. It included a detailed discussion of the dynamics of his “spatial proximity” model. Like Sakoda, Schelling’s model randomly arranged individuals and vacancies on a 2D lattice. Each individual was assigned a tolerance threshold of 0.50, that is individuals are unhappy when the neighborhood proportion of unlike neighbors exceeds 50 percent. Then, without particular order, individuals are selected for transfer to a tolerable vacant location. This procedure is repeated until all individuals are satisfied, or until no additional viable moves are available. The results were striking. Even without overt segregationist preferences, the collective dynamics of individuals’ preferences to avoid minority status resulted in near-total segregation.
Additionally, Schelling recognized that varying tolerances (preference schedules) and increasing neighborhood sizes could generate a greater variety of results, including reduced segregation, in his “bounded-neighborhood” model. In bounded-neighborhood models agents consider their membership in a discrete, finite neighborhood. This model was not, however, concerned with the properties of the configurations of individuals within the neighborhood. Rather, it only considered whether they chose to remain within or exit.
In the years after Schelling, substantial replication, extension, and discussion of the model took place \cite{clark2008understanding}. Despite Schelling’s acknowledgement of his model’s crudity, it gained importance in the 1980s and 1990s with debates concerning the genesis and perpetuation of segregation. On one side were proponents of the notion that segregation in the United States was primarily caused and perpetuated by discriminatory housing policies. The other side suggested that could be caused by the collective dynamics of ethnic preferences aided by economic disparities \cite{clark2008understanding}. The suggestion that segregation exists due to such collective dynamics has even been an important factor in U.S. court decisions as legal remedies to racial injustices often require evidence of de jure causes \cite{FrankenbergErica2018DFST}. As this controversy unfolded, several researchers examined and elaborated Schelling’s work.

\subsection{Empirically Varied Tolerance Schedules}
Among them, Clark \cite{ClarkW.A.V.1991RPaN} built upon Schelling’s “bounded-neighborhood” work by establishing tolerance schedules based on empirical data. Clark was able to provide validation for Schelling’s suggestion that heterogeneous tolerances could lead to a greater variety of results. However, Clark also predicted that mixed equilibria may be rarer than previously expected. It is important to note that this hypothesis applies to a single neighborhood without regard to its specific spatial configuration.
Later, Xie \& Zhou \cite{xie2012modeling} adopted the Detroit data used by Bruch \& Mare \cite{BRUCHElizabethE2006Ncan} taken from the Multi-City Study of Urban Inequality (MCSUI) to assign tolerance schedules to individuals based on a constructed Guttman scale \cite{GuttmanLouis1944ABfS}. Xie \& Zhou’s baseline model generated six classes of agents, five following the upper tolerance thresholds of each level of the Guttman scale and a sixth following a rank-ordered logit model. The rank-ordered logit model provides a probability that a candidate occupant will move to a specified neighborhood, thus incorporating an additional layer of stochasticity. Recognizing the unrealistic assumption of homogeneity within each level of the Guttman scale, Xie \& Zhou developed a continuous tolerance schedule by drawing from a uniform distribution bounded by each level’s lower and upper tolerance threshold. Using this tolerance schedule, Xie \& Zhou extended Schelling’s \cite{schelling1971dynamic} and Bruch \& Mare’s \cite{BRUCHElizabethE2006Ncan} work by conducting additional simulations. Their simulations employed transition rules in the same way as their predecessors but used the updated tolerance schedule. Their results demonstrated a substantial reduction in realized segregation resulting from the novel tolerance schedule.

\subsection{Parameter Elaboration}
Fossett \cite{fossett2006ethnic} extensively elaborated Schelling’s “spatial proximity” model to include a variety of additional parameter settings which enabled new experiments. Fossett established a random baseline model with a modified 2D lattice structure which produced no significant segregation. This structure left certain corner units of lattice subsections empty resulting in a rounded superstructure. Next, Fossett successively included additional agent parameters, including tolerances for status, ethnicity, and housing quality. Fossett also examined the impact of reducing agents’ perceptual ranges. Overall, Fossett’s work reinforced the view that micro-level interactions could generate and perpetuate segregation even in the absence of discrimination.

Abella, San Miguel, \& Ramasco \cite{AbellaDavid2022AeiS} incorporated an aging effect. Their model modifies an agent's probability of migration by making it inversely proportional to time spent in a satisfactory neighborhood. The authors examined several homogeneous tolerance regimes. The addition of the aging effect led to the surprising result that segregation was maintained even at higher tolerance levels.

Gambetta, Mauro, \& Pappalardo \cite{GambettaDaniele2023Mcis} introduced mobility constraints. In their model, preference for nearby or distant candidate vacancies was controlled by a distance exponent, $\beta$. For $\beta > 0$, more distant locations are preferred. For $\beta < 0$, nearby locations are preferred. In addition, they incorporated a ``relevance'' effect, controlled by a relevance exponent, $\alpha$, where $\alpha > 0$ indicates a preference for migration toward the center of the grid. These effects had significant impacts on behavior both independently and in concert. The distance model alone showed increased segregation for $\beta > 0$ and decreased segregation for $\beta < 0$. The relevance model alone showed modestly increased segregation for $\alpha > 0$. Increases in $\alpha$ were moderated when $\beta < 0$ as the preference for nearby locations counteracted the relevance effect.

Mantzaris \cite{MantzarisAlexanderV.2020Iamv} introduced a monetary variable  to address the Schelling model’s apparent decrease in entropy over time. By assigning agents incomes sampled from real-world data and linking residential moves to income dynamics, the model captures an overall trend toward higher entropy consistent with the second law of thermodynamics. Specifically, as agents move in search of residential homophily, they simultaneously adjust income distributions through expenditures, pushing system configurations closer to higher-entropy states.

Several other computational modeling efforts have examined the interplay between segregation and wealth inequality. Benard \& Willer showed that individual resource differences can enable segregation by status on a toroidal cellular automata model \cite{benard2007wealth}. Sahasranaman \& Jensen, in two papers, developed and extended a model of wealth-based segregation which well-approximated empirical evidence when individuals had a preference for neighborhoods in which their wealth was comparable to or greater than their neighbors \cite{sahasranaman2016dynamics,sahasranaman2017cooperative}. Gauvin et al. provide contradictory evidence \cite{gauvin2013modeling}. That is, their model which also provided a reasonable approximation of empirical data showed that wealth-based segregation could occur when individuals prefer to live in neighborhoods in which their wealth was less than or comparable to their neighbors. Harting \& Radi, used a model with heterogeneous tolerances and ethnic income disparities to show that when individuals make housing decisions based on economic factors in addition to ethnic preferences, income inequality may be a necessary condition for residential integration \cite{HartingPhilipp2020RsTr}. Bonakdar \& Roos used heterogeneous incomes and incorporated an auction process to determine housing prices based on demand \cite{Bonakdar2023}. Their simulation results supported the idea that when individuals have a preference for similar neighbors, population sorting motivates housing prices. Vicario implemented a model in which agents are characterized by a dynamic wealth variable showing that, absent perturbations at the individual level, wealth segregation emerges \cite{vicario2024dynamic}.

\subsection{Game-Theoretic Approaches}
Zhang \cite{zhang2004dynamic} translated Schelling’s “spatial proximity” model to a spatial game-theoretic model. In this, Zhang presumed asymmetrical tolerances between groups and added a simple housing market, again employing a 2D lattice. As prices in the housing market responded to demand, the asymmetry of tolerances played an important role. Housing for the group with more exclusionary tolerances became scarcer. Additionally, the quantity of housing units that were unsuitable for the exclusionary group began to exceed demand. The result was a significant disparity in prices based on neighborhood composition. Zhang showed that in the absence of such market influences, asymmetry of tolerances alone could not explain the genesis of segregation on the lattice.
In a separate paper, Zhang \cite{zhang2004residential} employed his game-theoretic approach on a 2D lattice to illustrate an extreme example: the genesis of segregation when all agents prefer total integration. This time, no vacancies were permitted on the lattice. Again, the asymmetry of tolerances was important. For each group, Zhang established a utility function with maximum payoff when there was a perfect mix of neighbors. Secondarily, when a perfect mix of neighbors was not available, agents preferred to avoid minority status. Zhang showed that this asymmetry of tolerances makes perfect mixing an unstable attractor. Whenever an agent finds a trading partner, the payoff gained by the former will always be outweighed by the penalty to the latter.
More recently, Grauwin et al. \cite{grauwin2012dynamic} built upon Zhang's work by employing evolutionary game theory to develop a potential function for analysis of Schelling's bounded-neighborhood model. Using this potential function, they analyzed stationary configurations generated by three different utility functions: linear utility, Schelling's original function, and asymmetric peaked utility functions. They showed that symmetric functions do not generate segregation, but, as in Zhang \cite{zhang2004residential}, any preference for majority status, even when individuals primarily prefer mixed environments, generates segregation. Their analytical tool enables exploration of an array of utility functions, including those rooted in empirical data.

\subsection{Heterogeneous Topology}

From 2015 to 2018, Gandica, Gargiulo, \& Carletti \cite{gandica2016can,GargiuloFloriana2015UsfS}  elaborated Schelling’s spatial proximity model using a metapopulation framework. Initially in \cite{gandica2016can}, they constructed regular 1- and 2-D lattices. Each node on the lattice could house a population of individuals with some limit $L$. Thus, for each individual living in the $i^{th}$ node, its neighborhood size is the level of population in $i$ plus the sum of the levels, $l$, of population in each neighboring node. The authors conducted simulations using several homogeneous tolerance thresholds, $\epsilon$. For $\epsilon\le0.5$, at the node level, total segregation was observed. 
It is important to note that the proposed metapopulation model is analogous to the instantiation of a heterogeneous topology superimposed on a lattice structure. This is equivalent to adding edges to a 1- or 2-D lattice populated by single occupant nodes to achieve the same heterogeneity of neighborhood sizes, except that the method proposed by Gandica, Gargiulo, \& Carletti increases the number of individual neighborhoods. 

Later Gandica, Gargiulo, \& Carletti \cite{GargiuloFloriana2015UsfS}, elaborated their earlier work on 2D lattices to examine the effect of varying network structures while employing the metapopulation model. While maintaining an average node-neighborhood size, k=4, small-world (Watts-Strogatz \cite{Watts1998}), random (Erdos-Renyi \cite{Erdos:1960}), and scale-free (Barabasi-Albert \cite{Barabasi.Albert.74.47}) networks were generated.  Given the same node-level population limit, L, a greater variety of neighborhood sizes is possible. Results from this inquiry showed no qualitative difference in the chosen metric for long-term average node tendency, asymptotic averaged node magnetization, $\langle u \rangle$, in Eq. 1. 
\begin{equation}
    \langle u \rangle = \frac{1}{N} \sum_{i} \frac{|n_i^B - n_i^A|}{n_i^B + n_i^A}
\end{equation}
where $n_i^A$ and $n_i^B$ are the number of A and B individuals in the $i^{th}$ node, respectively, and $N$ is the total number of nodes.  

Similarly, but without use of a metapopulation model, Fagiolo, Valente, \& Vriend \cite{FagioloGiorgio2007Sin} considered six distinct network topologies: 2D boundary-less lattices with von Neumann and Moore neighborhoods, regular and random non-directed graphs, and scale-free and small-world networks. The authors examined the impact of these topologies on the segregation process. Homogeneous tolerance thresholds were employed. Consistent with \cite{GargiuloFloriana2015UsfS}, the results showed that the segregation process was not significantly impacted by the network topology. In 2023, Su \& Zhang \cite{su2023significant} implemented Schelling's metapopulation model on a star-shaped network of $Q$ blocks. Each block with a population limit of $M$ agents was connected to a central hub. In addition, individual's migration was constrained to adjacent blocks. This construction substantially suppressed the segregation phase.

\section{Models}

\subsection{Baseline Model}

The baseline network model topology is a 32 by 32 regular lattice with edges connecting von Neumann neighborhoods (see Figure~\ref*{fig:fig_s2}) and a closed boundary condition.  As in Xie \& Zhou \cite{xie2012modeling}, approximately 15\% of nodes are reserved as excess housing.  The remaining nodes are randomly assigned either a red or blue occupant.  As in Schelling \cite{schelling1971dynamic}, each occupant is assigned an identical tolerance threshold, $\epsilon$. We set $\epsilon=0.3962$, indicating a tolerance for at most ~39.62\% opposite colored neighbors. To draw a more direct comparison, this $\epsilon$-value is obtained by taking the mean tolerance threshold from the populations of agents generated using tolerance schedules as described in the next section. 
Again, mirroring Xie \& Zhou’s \cite{xie2012modeling} baseline model, at time t, the current neighborhood proportion of dissimilar neighbors for the $i^{th}$ occupant in the $j^{th}$ neighborhood in $G$, $d_{ijt}$, is given as

\begin{equation}
    d_{ijt} = 
    \begin{cases} 
    \frac{n_{d_{i,t}}}{N_{jt}}, & \text{if } N_{jt} > 0 \\
    0, & \text{if } N_{jt} = 0 
    \end{cases}
\end{equation}
where $n_{d_it}$ is the number of dissimilar neighbors and $N_{jt}$ is the total number of neighbors in the $j^{th}$ neighborhood in $G$.  

At each time step $t$, $d_{ijt}$ is calculated for each individual $i$ in the set of occupants $O$ and the set of candidate occupants at time $t$,  $C_{Ot} = \left\{ i \mid i \in O, d_{ijt} > \epsilon_i \right\}
$, is constructed where $j$ is the individual’s current location and $\epsilon_i$ is its tolerance threshold. A candidate occupant $i\in C_{Ot}$ is randomly selected.  For all vacant nodes at time $t$, $v\in V_t$, neighborhood composition is calculated to create a list of candidate vacancies, $C_{Vt} = \{ j \mid j \in Vt, \ d_{ijt} \le \epsilon_i \}
$,  from which a destination node, $j\in C_{Vt}$ is randomly selected.   The candidate occupant at the $i^{th}$ node moves to the selected vacancy and leaves a vacancy in its place.  If at any time $t$, $C_{Ot}={\emptyset}$, all nodes are satisfied, and no additional trades will be found.  If at any time $t$, $C_{Vt}={\emptyset}$, the occupant at the selected node, $i\in C_{Ot}$, is unable to locate a satisfactory destination and remains in place.

\subsection{Heterogeneous Topology}

\begin{figure}[H]
    \centering
    \includegraphics[width=0.9\textwidth]{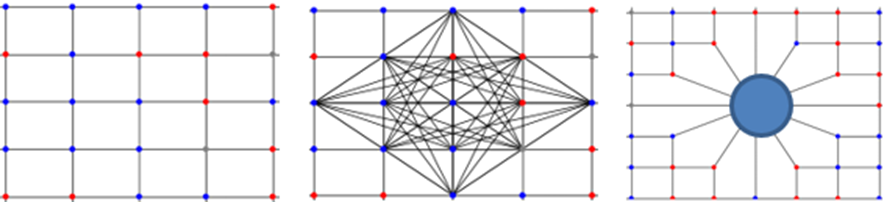}
    \caption{Left: an initial portion of 2D regular lattice with von Neumann neighborhoods; center: its densified counterpart; right: the metapopulation representation of the densification obtained by replacing the cluster with a single metapopulation node. Each node within the densified cluster has a link to all other nodes in the cluster. This is equivalent to their replacement with a single node as a container for the cluster.}
    \label{fig:fig_s2}
\end{figure}

Primarily to make topological differences more explicit, we depart from Gandica, Gargiulo, \& Carletti \cite{gandica2016can}, who implemented a metapopulation model, and instead select random neighborhoods to densify.  To do so, for each randomly selected node, a set of immediate neighbors, $N_0$, is constructed.  For each neighbor, $n_i \in N_0$, a set of second-order neighbors, $N_i\prime$ is constructed.  The union of these sets represents a cluster of nodes, $N_C$. Finally, a set of edges, $E={\left(n,m\right):n\in N_c,m\in N_c,\ n\neq m}$ is constructed. These edges are then added to G.  A single iteration of this procedure is considered a single densification as shown in Figure~\ref*{fig:fig_s2}.

This method ensures that neighborhood densities are distributed consistently across the network, so nodes and their neighbors cannot have unrealistic differences in degree.  The result of multiple densifications is substantially greater variation in node degree across the network as well as a marked increase in the mean node degree.  The result is a variety of neighborhood sizes. While in the base model, all neighborhoods are von Neumann neighborhoods (except at the boundary), randomly densified lattices have a variety of neighborhood boundary relationships, e.g., a von Neumann neighborhood where individuals have 4 neighbors, might be adjacent to a Moore neighborhood where individuals have 8 neighbors.  This enables a richer diversity of neighborhood relationships. 

\begin{figure}[H]
    \centering
    \includegraphics[width=0.5\textwidth]{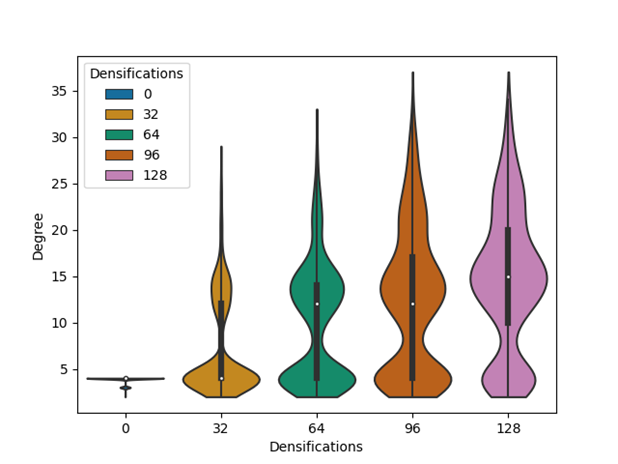}
    \caption{Impact of densifications on graph degree distribution. Distributions for 32, 64, 96, and 128 densifications are bimodal. There is an observable transition between dominant modes as the number of densifications increases.}
    \label{fig:fig_s4}
\end{figure}

As the number of densifications increases, substantial differences in degree distributions are observed, as shown in Figure~\ref*{fig:fig_s4} and Table~\ref*{tab:tab_1}. To ensure adequate topological heterogeneity is produced by the densification procedure, Table~\ref*{tab:tab_1} displays the associated Shannon entropies for the distributions of neighborhood sizes and neighborhood-size pair configurations at each level of densification.

 It is clear that the densification procedure effectively produces the desired heterogeneity of topology. The level of topological heterogeneity produced at 128 densifications exceeds the level produced by any network generator in \cite{gandica2016can}. Because there is an upper bound on local density, successive increases in densifications lead to progressively smaller increases in heterogeneity of neighborhood size and neighborhood-size pair configurations. Eventually, continued densification returns the topology to homogeneity. Here, we ensure our densification procedure does not inadvertently produce a homogeneous topology by limiting the number of densifications to 128.

\begin{table}[H]
    \centering
    \begin{tabular}{c|ccccccc}
    \hline
    D & $\mu$ & $\sigma$ & $\eta$ & H\(_k\) & H\(_{pairs}\) & \( H(T)_0 \) & \( H(T)_F \) \\
    \hline
    0   & 3.875  & 0.342  & 4   & .557  & .682  & 7.651 & 5.842 \\
    32  & 7.301  & 5.266  & 4  & 2.300 & 4.878 & 7.646 & 6.155 \\
    64  & 10.148 & 6.326  & 12 & 3.218 & 6.312 & 7.654 & 6.138 \\
    96  & 12.477 & 7.579  & 12 & 3.805 & 7.164 & 7.636 & 6.163 \\
    128 & 14.670 & 7.590  & 15 & 4.190 & 7.642 & 7.642 & 6.254 \\
    \hline
    \end{tabular}
    \caption{Mean, standard deviation, median degree, entropy of node degree, entropy of neighborhood-size pairs, and initial and final entropies for the distribution of pairs' tolerances.}
    \label{tab:tab_1}
\end{table}

\subsection{Heterogeneous Tolerances}
\begin{figure}[H]
    \centering
    \includegraphics[width=0.5\textwidth]{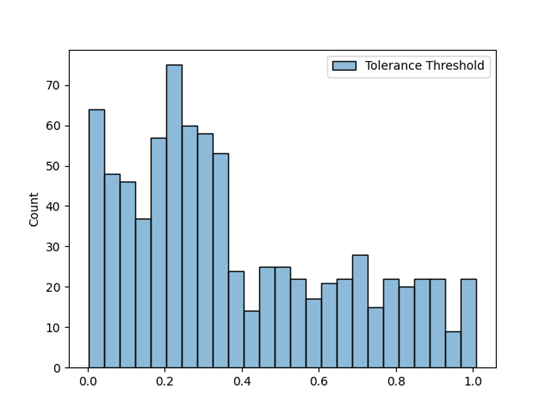}
    \caption{Histogram of Guttman scale tolerance thresholds initialized for a single simulation instance. Not shown: 4.25\% of individuals obeying Xie \& Zhou’s rank-ordered logit model transition function.}
    \label{fig:fig_s3}
\end{figure}

Following Xie \& Zhou \cite{xie2012modeling}, agents are provided with heterogeneous tolerances (tolerance thresholds) aligned with the Guttman scale and rank-ordered logit model derived from Bruch \& Mare’s \cite{BRUCHElizabethE2006Ncan} Detroit data. Tolerance thresholds were assigned by drawing values from a uniform distribution over a given interval: For 10.47\% of individuals, $\epsilon_i$ fell within $[0.0,0.07)$; for 18.10\% of individuals, $\epsilon_i$ fell within $[0.07,0.21)$; for 26.73\% of individuals, $\epsilon_i$ fell within $[0.21, 0.36)$; for 13.86\% of individuals $\epsilon_i$ fell within $[0.36,0.57)$; for 26.59\% of individuals, $\epsilon_i$ fell within $[0.57,1.00]$.  For these individuals, the simulation procedure described in the previous section was implemented with $\epsilon_i$ replacing $\epsilon$. Figure~\ref{fig:fig_s3} shows an example distribution of tolerance thresholds drawn from the constructed distribution. For the 4.25\% of individuals in the Detroit data found not to conform to the Guttman scale, no static tolerance threshold was set. Instead, Xie \& Zhou’s rank-ordered logit model (Eq. 4 in \cite{xie2012modeling}) was implemented to determine the probability of transition to each candidate neighborhood.  The transition destination, $j\in C_{Vt}$, is then randomly selected with probability, $\widehat{p_{jt}}$, given by the model:
\begin{equation}
    \hat{p}_{ijt}=\frac{\exp{\left(13.0d_{ijt}-17.9d_{ijt}^2\right)}}{\sum_{k \in C_{Vt}}{\exp\left(13.0d_{ikt}-17.9d_{ikt}^2\right)}}
\end{equation}

Here, ${\hat{p}}_{ijt}$, represents the normalized probability that the $i^{th}$ occupant will move to the $j^{th}$ neighborhood in the set of candidate vacancies. The models coefficients were derived by Xie \& Zhou via maximum likelihood estimation from Bruch \& Mare's Detroit data. This model effectively weights the probability of a move to each candidate vacancy by its proximity to an estimated central tolerance threshold, $\epsilon\approx.3631$, on the cusp of the third and fourth intervals noted above.  

The adoption of Xie \& Zhou’s tolerance model increases the Shannon entropy of tolerance thresholds, $H_\epsilon$, from $0$ in the homogeneous Schelling case to at least $H_\epsilon\approx5.102$. This approximation was calculated by discretizing assigned tolerance thresholds into 37 identically sized bins and allowing non-Guttman scale agents to occupy an additional distinct bin.

\section{Experiments \& Results}
For the primary set of experiments, ten unique parameter configurations were used: Half employed uniform Schelling \cite{schelling1971dynamic} tolerance threshold assignments and half used heterogenous Xie \& Zhou \cite{xie2012modeling} tolerance threshold assignments. For each tolerance assignment method, simulations were implemented in Python and conducted with 0, 32, 64, 96, and 128 densifications. For each parameter setting, 100 simulations were run for 4000 time steps. Convergence was defined as the cumulative change in assortativity remaining below 0.01 over any 100-step rolling window. All simulations met this criterion within the allotted timeframe. To ensure robustness of results, additional simulations were conducted with increased vacancy on the lattice and with the presence of a minority population.

\subsection{Segregation}
Newman’s \cite{PhysRevE.67.026126} assortativity coefficient, $r$, a well-established measure of categorical mixing on networks, is used as a measure of segregation levels during the simulations. $r$ is given by Eq. 2 in \cite{PhysRevE.67.026126}:
\begin{equation}
    r=\frac{\sum_{i} e_{ii}-\sum_{i}{a_ib_i}}{1-\sum_{i}{a_ib_i}}
\end{equation}
where $e_{ij}$ denotes the fraction of edges which connect a node of type $i$ to one of type $j$, $a_i=\sum_{j} e_{ij}$, and $b_j=\sum_{i} e_{ij}$. Thus, for a graph with occupants with binary attributes, we can write $a_i=b_i$ and $e_{ii}=1-a_i$. When node color populations are approximately equal, we may make the following approximation with $\sum_{i}{a_ib_i}\approx.5$:
\begin{equation}
    r = \frac{\sum_{i} e_{ii}-\sum_{i}{a_ib_i}}{1-\sum_{i}{a_ib_i}}\approx\frac{1-e_{ij}-.5}{1-.5}=1-2e_{ij}
\end{equation}
with $e_{ij}$ denoting the fraction of edges connecting dissimilar neighbors. Thus, a graph with exactly half of its edges connecting dissimilar neighbors would have $r=.5$, while graphs with all or none of their edges connecting dissimilar neighbors would have $r=0$ and $r=1$, respectively. Values of $r$ close to 1 indicate high levels of segregation while values of $r$ close to 0 indicate approximately random mixing.  To account for vacant nodes on the network, the assortativity coefficient for the subgraph containing only occupied nodes is used.  For all simulations where the derived subgraph consisted of multiple connected components, $r$ was calculated for each connected component and a weighted average was constructed:
\begin{equation}
    r^\prime=\frac{1}{N}\sum r_in_i
\end{equation}
where $N$ is the total number of occupied nodes and $r_i$ and $n_i$ are the assortativity coefficient and size of the $i^{th}$ component, respectively. For connected components with homogeneous composition, a value of 1 was assigned for $r$. To observe the process of segregation, $r^\prime$ was recorded at each time step.

\begin{figure}[H]
    \centering
    \includegraphics[width=.9\textwidth]{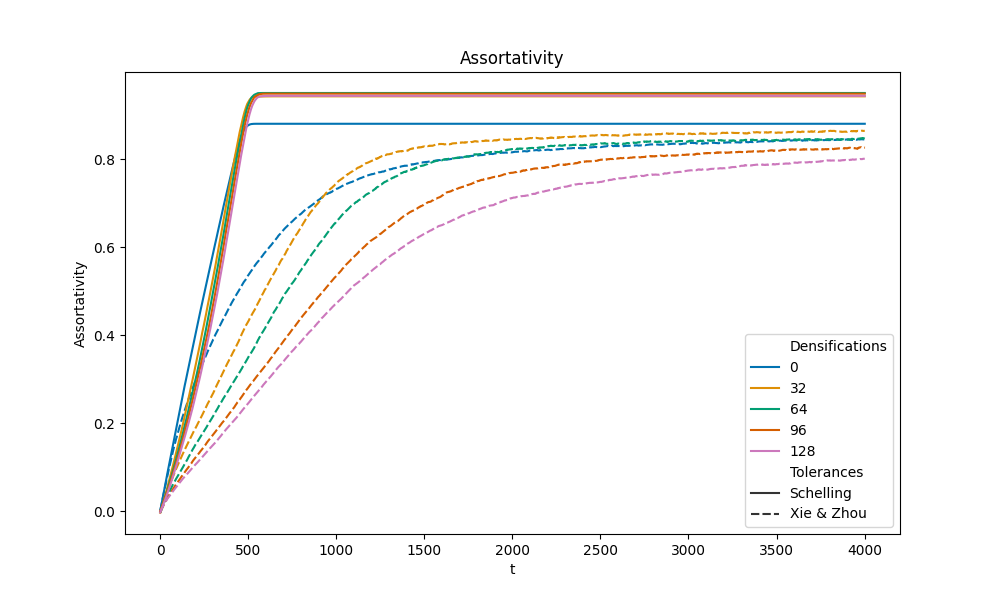}
    \caption{Average assortativity over time for each group of simulations.}
    \label{fig:assort}
\end{figure}

\begin{figure}[H]
    \centering
    \includegraphics[width=0.9\textwidth]{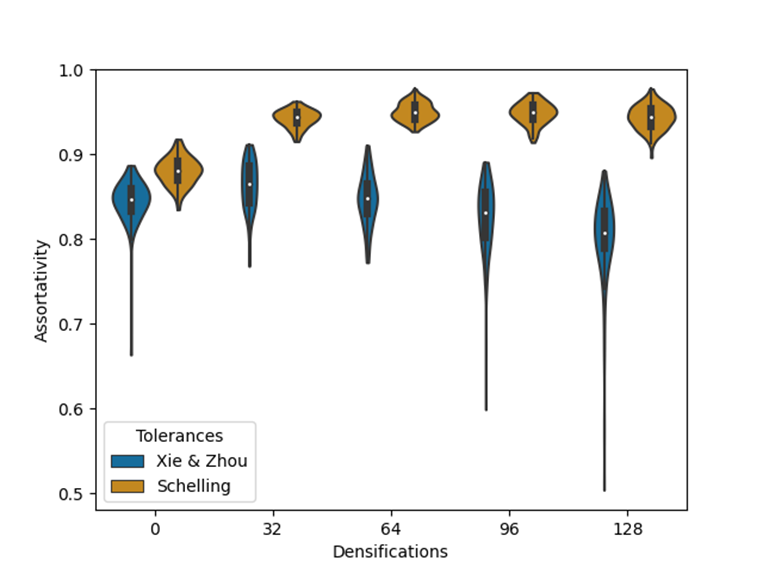}
    \caption{Final graph assortativity for each group of simulations.}
    \label{fig:fig1}
\end{figure}

The observed mean final assortativity was higher for all Schelling-tolerance simulation runs. As in Xie \& Zhou \cite{xie2012modeling}, heterogeneity of tolerances did, on their own, result in reduced assortativity.  Consistent with Gandica, Gargiulo, \& Carletti’s \cite{gandica2016can} observations, increasing topological heterogeneity alone did not produce results with reduced assortativity when $\epsilon\approx0.39$, in fact, any increase in topological heterogeneity for Schelling-tolerance simulations resulted in a marked increase in realized assortativity. The combination of both dimensions of heterogeneity resulted in progressive reductions in assortativity. Xie \& Zhou-tolerance simulations generated assortativity more slowly and resulted in reduced levels of segregation compared with Schelling-tolerance simulations at every level of densification. The lowest levels of assortativity observed occurred with Xie \& Zhou \cite{xie2012modeling} tolerances and 128 densifications. Results are illustrated in Figures~\ref*{fig:assort} and \ref*{fig:fig1}.

The introduction of heterogeneity of topology had the effect of increasing assortativity in all Schelling-tolerance simulation runs due to the presence of densified clusters that increased the number of edges connecting nodes with similar occupants. These increases in assortativity began to disappear when heterogeneities were combined, especially when significant topological heterogeneity was present. 

To ensure robustness, additional simulations were conducted with 30\% excess housing on the lattice. This modification moderated reductions in assortativity previously observed as a result of densification, however, heterogeneity of tolerances persisted in reducing assortativity. This is an indication that abundant excess housing enables individuals to avoid the influence of densified clusters. 

Curiously, when the presence of a minority population was modeled by assigning 30/70 red/blue occupants, Schelling tolerances yielded greater reductions in assortativity than Xie \& Zhou tolerances. Visual inspection of the final graphs revealed the tendency, in the Schelling case, for individuals in the minority population to become isolated and trapped, surrounded by dissimilar neighbors but without any suitable candidate vacancies. This effect was less pronounced when Xie \& Zhou tolerances were assigned.

\subsection{Organization}
Shannon entropy, $H$, of the distribution of the tolerance levels for each connected pair was observed for the population of graphs both at initialization and at the $4000^{th}$ time step. Since tolerances are drawn from a continuous distribution, tolerance levels are discretized within 25 equal-sized partitions to calculate the Shannon entropy.  Vacancies and non-Guttman tolerance individuals are each assigned to monolithic bins. More explicitly,
\begin{equation}
    H\left(T\right)=\ -\sum_{x \in X}{P\left((t_{ij})\right)\log_2{P((t_{ij}))}}
\end{equation}
where $t_{ij}$ is the pair of discretized tolerances, $\left(t_i,t_j\right)|\ i,\ j\in G$, observed.
	
As another indicator of organization, final mean degree was observed for various node types for comparison with the final mean graph degree. Mean degree was observed for vacancies, occupied nodes, nodes occupied by highly tolerant individuals $(\epsilon\geq0.57)$, and nodes occupied by highly intolerant individuals $(\epsilon\le0.07)$.

\subsubsection{Hub Composition}
\begin{figure}[H]
    \centering
    \includegraphics[width=0.9\textwidth]{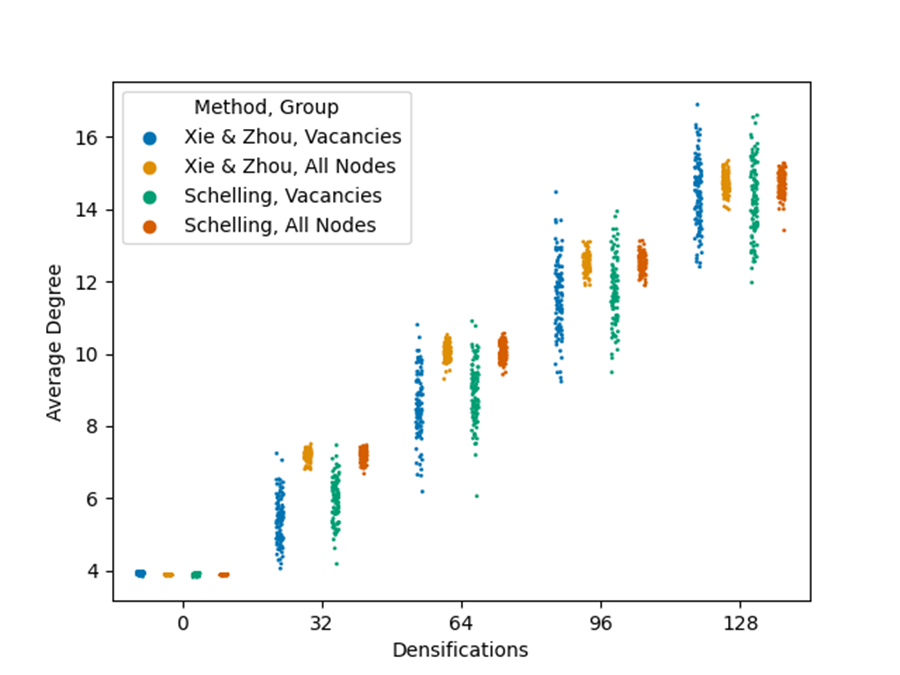}
    \caption{Final mean vacancy and node degree for each simulation group.}
    \label{fig:fig2}
\end{figure}

The mean degree for each graph and its vacancies was calculated; see Figure~\ref*{fig:fig2}. The mean vacancy degree showed greater variability than the graph degree for each group.  No significant difference was observed between the accumulations of vacancies in Xie \& Zhou tolerance simulations and Schelling tolerance simulations.  When substantial levels of densification were present, the mean vacancy degree could exceed the mean graph degree. Otherwise, the mean vacancy degree tended lower. 
 
\begin{figure}[H]
    \centering
    \includegraphics[width=1\textwidth]{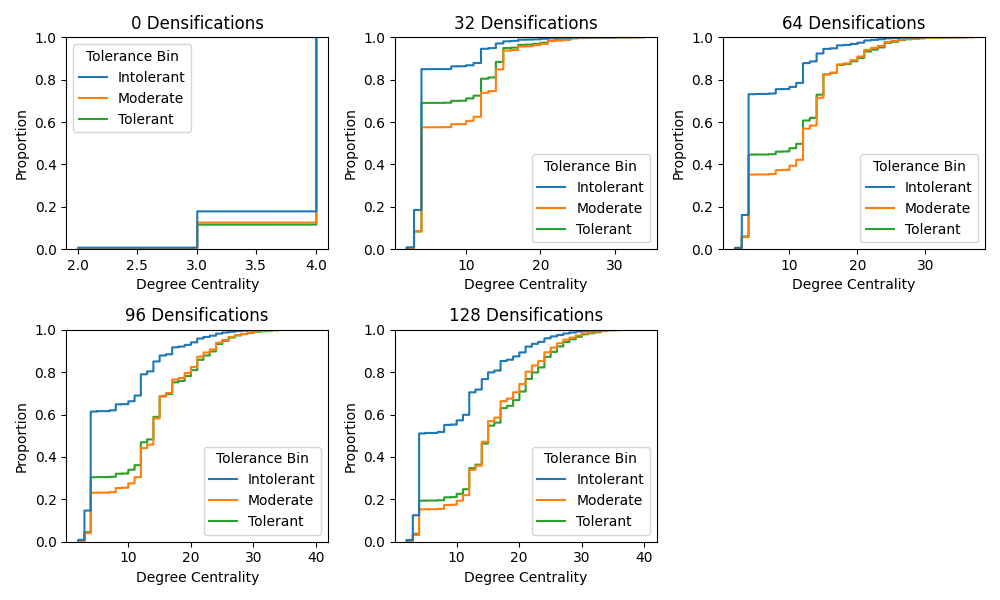}
    \caption{Empirical cumulative distribution functions (ECDFs) of degree centrality for tolerant, moderate, and intolerant individuals across different densification levels. The plots illustrate the increasing divergence in centrality distributions as network density increases.}
    \label{fig:fig3}
\end{figure}

\begin{figure}[H]
    \centering
    \includegraphics[width=0.9\textwidth]{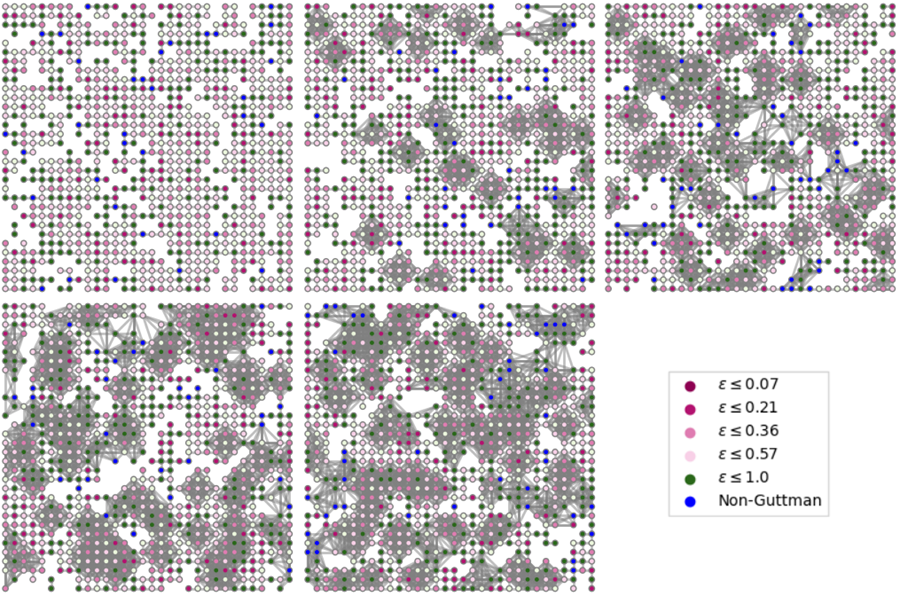}
    \caption{Sample final graph images for 0, 32, 64, 96, and 128 densifications using Xie \& Zhou tolerances. Redder colorings represent less tolerant individuals, greener colorings represent more tolerant individuals. Blue coloring represents non-Guttman individuals. Vacancies are removed.}
    \label{fig:fig4}
\end{figure}

For simulations with Xie \& Zhou tolerances on densified graphs, on average, individuals in the group with the highest tolerance thresholds ($\geq0.57$) had distinctly higher degree centralities than those in the group with the lowest tolerance thresholds ($\le0.07$). This difference became more pronounced as the number of densifications increased. These results are illustrated in Figures~\ref*{fig:fig3}--\ref*{fig:fig4}. To confirm the relationship between tolerance level and position within networks of varying density, we used 2-sample Kolmogorov-Smirnov (K-S) tests comparing the degree centrality distributions of tolerant ($\epsilon \geq .57$), moderate ($.07 \leq \epsilon < .57$), and intolerant ($\epsilon < .07$) individuals across five levels of graph densification \cite{Massey1951TheKT}. The K-S $D$-statistic, which captures the maximum difference between empirical cumulative distribution functions, revealed that as network density increased, so did the divergence in centrality between tolerance groups. For example, at low density, the largest $D$-statistic was $D = 0.063$, $p = 1.08 \times 10^{-22}$ comparing intolerant individuals with moderates, but by the highest densification level, the largest gap, between tolerant and intolerant individuals, had widened substantially ($D = 0.379$, $p < 10^{-300}$). The $D$-statistic comparing intolerant with moderate individuals increased monotonically with densification. The $D$-statistic comparing intolerant with tolerant individuals peaked at 96 densifications. The $D$-statistic comparing moderate with tolerant individuals peaked at 32 densifications. These results suggest that in more interconnected networks, tolerance becomes increasingly predictive of the topological distribution of individuals. From these results, we conclude that both tolerant and moderate individuals are more likely to occupy high-degree nodes in denser networks, while intolerant individuals are more likely to occupy low-degree nodes. 

\begin{table}[ht]
\centering
\begin{tabular}{|c|c|c|c|c|}
\hline
\textbf{Densification} & \textbf{Comparison} & \textbf{$D$-statistic} & \textbf{$p$-value} & \textbf{Centrality with max diff} \\
\hline
0   & Intolerant vs Moderate & 0.053 & $1.58 \times 10^{-19}$ & 3 \\
    & Intolerant vs Tolerant & 0.063 & $1.08 \times 10^{-22}$ & 3 \\
    & Moderate vs Tolerant   & 0.009 & 0.119 & 3 \\
\hline
32  & Intolerant vs Moderate & 0.274 & $<10^{-300}$ & 4 \\
    & Intolerant vs Tolerant & 0.163 & $3.32 \times 10^{-152}$ & 8 \\
    & Moderate vs Tolerant   & 0.115 & $6.65 \times 10^{-182}$ & 4 \\
\hline
64  & Intolerant vs Moderate & 0.382 & $<10^{-300}$ & 8 \\
    & Intolerant vs Tolerant & 0.295 & $<10^{-300}$ & 8\\
    & Moderate vs Tolerant   & 0.094 & $4.32 \times 10^{-122}$ & 4 \\
\hline
96  & Intolerant vs Moderate & 0.395 & $<10^{-300}$ & 8 \\
    & Intolerant vs Tolerant & 0.329 & $<10^{-300}$ & 11 \\
    & Moderate vs Tolerant   & 0.073 & $5.98 \times 10^{-74}$ & 4 \\
\hline
128 & Intolerant vs Moderate & 0.379 & $<10^{-300}$ & 11 \\
    & Intolerant vs Tolerant & 0.358 & $<10^{-300}$ & 12 \\
    & Moderate vs Tolerant   & 0.041 & $3.77 \times 10^{-23}$ & 4 \\
\hline
\end{tabular}
\caption{K-S Test Results by Densification Level: Centrality Differences Among Tolerance Groups}
\label{tab:ks_densification}
\end{table}

\subsubsection{Paired Tolerances}
\begin{figure}[H]
    \centering
    \includegraphics[width=.9\textwidth]{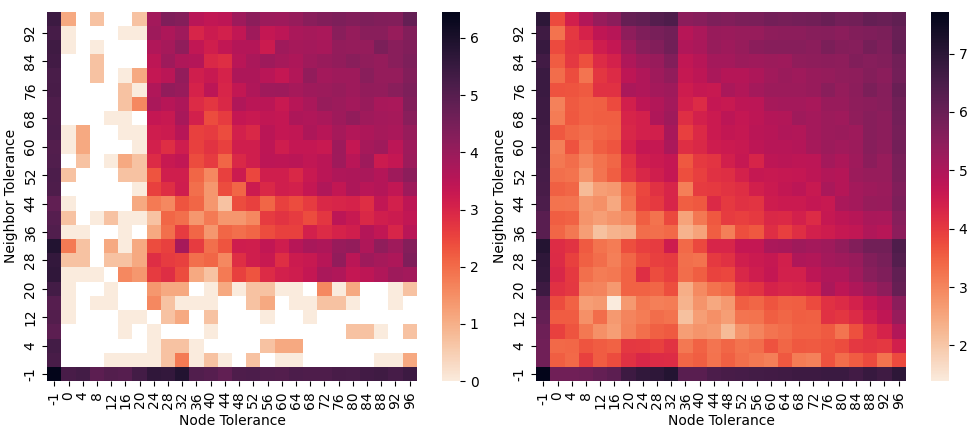}
    \caption{Heatmaps show the log frequencies of tolerance-level pairs for dissimilar neighbors. Left, 0 densifications; right, 128 densifications. A tolerance level of -1 denotes non-Guttman individuals.}
    \label{fig:fig5}
\end{figure}

In addition to the topological organization of tolerances, the pairwise organization of tolerances can also be observed. At every level of densification, the distribution of pairs’ tolerances became organized as segregation increased. This effect became less pronounced as the number of densifications increased.  At all levels of densification, distinct regions of like-tolerance nodes can be observed. This organization is further illustrated in Figure~\ref*{fig:fig5} where the relative frequencies of tolerance-pairs for dissimilar neighbors are shown. The overall level of organization exhibited at each level, the difference in levels of heterogeneity, is shown in Table~\ref*{tab:tab_1}. As densifications increase, final graphs are less organized.

This result was robust to the presence of a minority population, but an increase in excess housing increased the organization of tolerance pairs by enabling intolerant individuals to avoid contact. Figure~\ref{fig:fig5} shows the organization of tolerance pairs for dissimilar neighbors at 128 densifications when the excess housing rate is 30\%.

\begin{figure}[H]
    \centering
    \includegraphics[width=.7\textwidth]{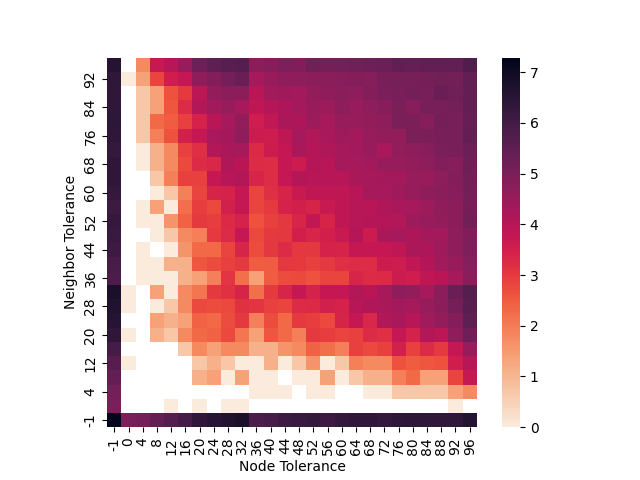}
    \caption{This heatmap shows the log frequencies of tolerance-level pairs for dissimilar neighbors when the excess housing rate is 30\%. A tolerance level of -1 denotes non-Guttman individuals.}
    \label{fig:fighm}
\end{figure}

\section{Discussion}
The results of this study show that the introduction of multiple dimensions of heterogeneity in network models of residential segregation can lead to significant changes in the dynamics of segregation. The combination of topological and behavioral heterogeneities produces reductions in segregation beyond those observed with either dimension alone. In addition, the presence of these complementary heterogeneities lead to the emergence of two notable dynamics: ordered migration and tolerance repelling intolerance.

\subsection{Ordered Migration}
\begin{figure}[H]
    \centering
    \includegraphics[width=0.9\textwidth]{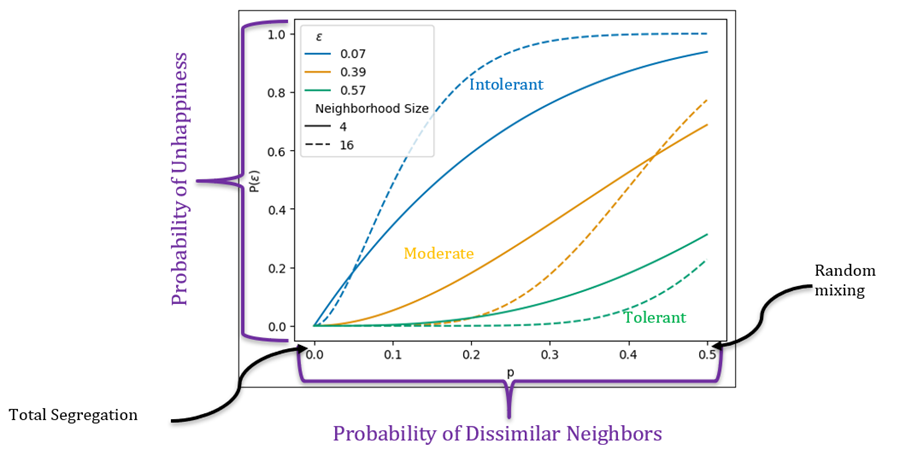}
    \caption{Vertical axis: Probability of exceeding tolerance threshold, $P_\epsilon$. Horizontal axis: probability of $n^{th}$ dissimilar neighbor, $p$. Note that as assortativity increases, the probability of dissimilar neighbors decreases.}
    \label{fig:fig6}
\end{figure}

The probability that an individual will exceed its tolerance threshold, $P_{d_{it}>\epsilon}$, can be given as a function of the probability, p, that at least k neighbors will be dissimilar:
\begin{equation}
    P_{d_{ij}>\epsilon}(p)=\sum_{k=\left\lceil\epsilon\ast N\right\rceil}^{N}\left(\binom{N}{k}p^k\left(1-p\right)^{N-k}\right)
\end{equation}
where $\epsilon$ is the individual’s tolerance threshold and N is the neighborhood size. Since our models begin with a randomly mixed population where $r\approx0$ and $r$ increases approximately monotonically, we need only consider cases where $0\le p\le0.5$, as shown in Figure~\ref*{fig:fig6}. As the migration process unfolds, $p$ decreases. For $\epsilon=0.39$, with only heterogeneity of topology, individuals in smaller neighborhoods will generally become more likely to exceed their tolerance thresholds than those in larger neighborhoods as assortativity increases over time. In this way, an additional dynamic is included: the out-migration of non-dominant-type individuals in densified clusters will precede in-migration to those clusters by dominant-type individuals. This behavior enables the increase in assortativity observed in Schelling-tolerance simulations with topological heterogeneity.
When both topological and behavioral heterogeneities are represented, dynamics are richer. The probability of an individual’s residence in a location that exceeds its tolerance threshold can be roughly ordered from greatest to least: low tolerance, large neighborhood; low tolerance, small neighborhood; moderate tolerance, small neighborhood; moderate tolerance, large neighborhood; high tolerance, small neighborhood; and high tolerance, large neighborhood. This ordering holds for $0.1\le p\le0.4$. While this model assumes independent and identically distributed neighbor types—an assumption that holds approximately only at initialization—its predictions serve as a theoretical baseline, offering valid insight into early-stage dynamics and comparative sensitivity to dissimilarity under varying tolerance thresholds and neighborhood sizes.

\subsection{Tolerance Repels Intolerance}
\begin{figure}[H]
    \centering
    \includegraphics[width=0.9\textwidth]{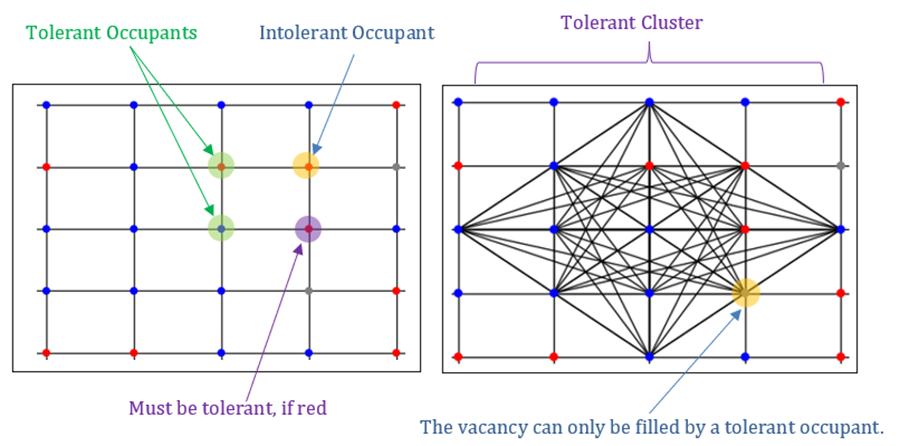}
    \caption{Left, a pair of highly tolerant dissimilar neighbors (highlighted in green). The upper node in the pair has an intolerant, similar neighbor (highlighted in orange). Given this configuration, for a red node to occupy the space highlighted in purple, it must possess a high tolerance threshold. Right, a group of tolerant dissimilar neighbors inside a densified cluster. The upper node in the pair has a vacant neighbor (highlighted in orange). Given this configuration, an intolerant node will likely never occupy the space highlighted in orange, regardless of its color. The tolerant nodes are highly entrenched.}
    \label{fig:fig7}
\end{figure}

When heterogeneity of tolerances is represented, pairs of dissimilar individuals with high tolerance thresholds create another interesting dynamic: these nodes repel intolerant nodes and attract tolerant neighbors. This dynamic is restrained on a 2D grid with von Neumann neighborhoods. Such pairs can have no common neighbors, but some neighbors of each will share an edge; see Figure~\ref*{fig:fig7}. With network topologies where dissimilar adjacent nodes may share a neighbor, that neighbor must possess sufficient tolerances to accommodate at least one pair of dissimilar neighbors.

When both heterogeneous tolerances and heterogeneous topology are represented, the influence of tolerant nodes is amplified by densifications. A densified cluster populated by dissimilar, tolerant nodes acts as a strong repellant for intolerant nodes as tolerant pairs will directly share more neighbors. Second, a densified cluster occupied by similar nodes may continue to attract similar nodes but can no longer repel a dissimilar individual with a sufficiently large tolerance threshold. The introduction of relatively few highly tolerant dissimilar individuals to such a cluster will motivate catastrophic out-migration of intolerant individuals. These dynamics lead some densified clusters to become stores of tolerance and diversity. As the level of densification increases, the presence of tolerant nodes and densified clusters makes extremely homogeneous areas rarer. As this occurs, the most intolerant nodes can no longer find suitably homogeneous candidate vacancies and increasingly remain dissatisfied and connected to tolerant, dissimilar nodes; see Figure~\ref{fig:fig5}. Thus, both the organization of tolerances and the organization of color-types in the final networks are reduced; see Table~\ref*{tab:tab_1}. It is important to note, however, that the presence of excess housing can moderate these dynamics by providing intolerant individuals with more opportunities to avoid dissimilar neighbors.

Our models did not incorporate random movements by otherwise satisfied individuals. This likely moderated tolerant individuals' ability to monopolize densified clusters.  The presence of such random movements would likely lead to a more pronounced effect of tolerance repelling intolerance. This is a critical avenue for future work. Rather, to the extent that it is observed, tolerant individuals' dominance of densified clustered is an example of sensitivity to initial conditions. The aforementioned entrenchment of tolerant individuals in densified clusters at initialization is, on its own, sufficient to produce this dynamic. 

\section{Conclusion and Future Work}
In this study, we investigated the impact of representing multiple dimensions of heterogeneity in network models of residential segregation. Models which combine heterogeneity of tolerances with heterogeneity of topologies showed substantially different behavior than those which employed only one dimension of heterogeneity.  Ordered migration and the resultant clusters of tolerance break down intolerant homogeny. This has bearing on the important question raised previously: to what extent can the collective dynamics of individual preferences lead to residential segregation?  These results indicate that these collective dynamics may not contribute to residential segregation as much as previously thought if sufficient heterogeneity is represented.  At a minimum, these results indicate the necessity of representing both heterogeneity of tolerances and heterogeneity of topologies in network models of residential segregation, as the omission of one or the other will result in the loss of these significant model behaviors.  

It is notable that the dynamics which break down segregation along racial lines appear to create segregation in another way: the most tolerant individuals can become separated from the least tolerant individuals. This finding, paired with the results in Sayama \& Yamanoi \cite{sayama2020beyond}, may explain ongoing cultural fragmentation between urban and rural areas. The heterogeneity of tolerances and topology necessary for simultaneous maintenance of cultural diversity and social cohesion manifests segregation between tolerance levels themselves. This segregation ensures that opportunities for cultural diffusion between the intolerant subsets of population will be scarce thereby maintaining cultural diversity within the network. Here, rather than a fully connected network with weighted cultural diffusion rates, the network self-organizes such that the topology moderates cultural diffusion. This dynamic occurs even in the absence of agents’ preferences to occupy central locations on the network such as in Gambetta, Mauro, \& Pappalardo's \cite{GambettaDaniele2023Mcis} relevance model. We hypothesize that the addition of such a secondary preference would lead to a more dramatic level of segregation between tolerance levels since highly tolerant individuals who are not motivated by dissimilar neighbors would primarily be driven by this secondary preference. In such a scenario, highly tolerant individuals would be driven to dense areas and push intolerant individuals to sparse areas. Further work should investigate the influence of spatial relevance on the dynamics of segregation in networks with multiple dimensions of heterogeneity. 

This work has implications for social and organizational cultural diffusion. The results suggest that the organization of tolerance levels in a network can be influenced by the network’s topology. As such, the organization of cultural diffusion in a network may be influenced by the network’s topology. This effect would be most pronounced in social and organizations systems where segregation along cultural lines is likely to occur. Future work should investigate the impact of network topology on cultural diffusion in social and organizational systems. Additionally, this work did not consider the effect of economic or other social factors on the dynamics of segregation. In Zhang, a notable result arising from asymmetric preferences was that individual who prefer homogeneity seemed to congregate \cite{zhang2004dynamic}. Our result is consistent with this observation, however the mechanism is different, and the congregation is observed in sparse areas. Where Zhang employed simulated economic incentives, our result shows that topological heterogeneity can motivate aggregation of intolerant individuals as well. Future work incorporating these factors may provide additional insight into the dynamics of segregation in networks with multiple dimensions of heterogeneity.

This research may hold importance for policymakers. The results suggest that heterogeneity of topology can either moderate or amplify segregation depending on the proportion and density of available excess housing. If policymakers aim to reduce segregation in communities, they may consider replacing low-density housing with high-density housing. Simply continuing to add both types of housing stock will continue to provide opportunities for tolerant individuals to self-organize in high-density areas while intolerant individuals continue to retreat to low-density areas. Given the variability and flexibility of zoning ordinances in the United States, this may never be a feasible policy option. A comparative analysis of real-world mixing patterns in high-density and low-density areas would be a valuable next step in this research. Another approach may examine the proportions of low- and high-density housing included in municipalities' development plans. Emphasis on low-density housing may be a predictor of future segregation and urban-rural cultural fragmentation while underemphasis on low-density housing may predict a decline in cultural diversity.

\section{Acknowledgement}
A Thesis has previously been published \cite{DeterWill2023BaTH}.

\bibliography{bibliography}

\end{document}